\documentclass[
    ,final            
  ]
  {aipproc}

\layoutstyle{6x9}

\begin{document}

\title[Azimuthal anisotropy of $K^0_S$ and $\Lambda$ production]{ Azimuthal
anisotropy of $K_S^0$ and $\Lambda$ production at mid-rapidity from Au+Au
collisions at $\sqrt{s_{_{NN}}}=130$~GeV }

\author{Paul Sorensen for the STAR Collaboration}{ address={University
of California, Los Angeles, California 90095} }

\begin{abstract}   We report STAR results on the azimuthal anisotropy
parameter $v_2$ for strange particles $K_S^0$ and $\Lambda$ at
mid-rapidity in Au+Au collisions at $\sqrt{s_{_{NN}}}=130$~GeV at
RHIC.  The value of $v_2$ as a function of transverse momentum $p_t$
and collision centrality is presented for both particles and compared
with model calculations.  A strong $p_t$ dependence in $v_2$ is
observed up to $p_t \sim 2.0$~GeV/c where $v_2$ begins to saturate.
\end{abstract}

\maketitle

\section{Introduction}
Measurements of azimuthal anisotropies in the transverse momentum
distributions of particles probe early stages of ultra-relativistic
heavy-ion collisions \cite{sorge99,sorge97,ollitrault92}.  After an
initial geometric anisotropy is established in an off-center or
non-central collision, rescattering in the overlapping region of the
colliding nuclei amongst collision participants transfers the spatial
anisotropy into an anisotropy in momentum space.  The extent of the
transformation depends on the initial conditions and the dynamical
evolution of the collision. As a result, anisotropy measurements for
nucleus-nucleus collisions at RHIC energies may increase our
understanding of the processes governing the evolution of the
collision system and in particular, may provide information about an
early partonic stage in the evolution of the system
\cite{sorge99,gyulassy99,pasi01,gyulassy01,shuryak01,zlin01}.

For the purpose of studying azimuthal anisotropies it is advantageous
to write the triple differential distribution of particles in the form
of a Fourier series
\begin{equation}
E\frac{d^3N}{d^3p} = \frac{1}{2\pi}\frac{d^2N}{p_tdp_tdy}(1 +
\sum_{n=1}^{\infty}{2v_n cos(n\phi)}),
\end{equation}
where $p_t$ is the transverse momentum of the particle, $y$ is its
rapadity and $\phi$ denotes its azimuthal angle of emmision with respect to
the true reaction plane angle\footnote{The reaction plane is the plane
defined by the beam line and the line connecting the centers of the
colliding nuclei.}~\cite{yingchao96,art98}.  The harmonic
coefficients~$v_n$ are anisotropy parameters and the second
coefficient $v_2$ is called the {\it elliptic flow} parameter. Recent
experimental results from
RHIC~\cite{starflow1,starflow2,phenixv2qm01,phobosv2qm01} include
measurements of $v_2$ as a function of collision centrality and $p_t$
for charged particles up to a $p_t$ of about 2.0~GeV/c, and for
identified $\pi^{\pm}$, $K^{\pm}$ and $p(\overline{p})$ up to a $p_t$ of
about 0.8~GeV/c.

We report on the first measurement of the azimuthal anisotropy
parameter~$v_2$, as a function of $p_t$ and collision centrality, for
the strange particles $K_S^0$ and $\Lambda$ from minimum bias Au + Au
collisions at $\sqrt{s_{_{NN}}}=130$~GeV. Our measurements of $v_2$
for identified particles using the Solenoidal Tracker At RHIC (STAR)
are the first to extend beyond the $p_t$ range where particles are
identified by their specific energy loss~(dE/dx) in the gas of the
Time Projection Chamber~(TPC), and up to $p_t \sim 3.0$~GeV/c.
Previously $v_2$ in this higher $p_t$ range had only been measured for
unidentified charged particles~\cite{qm2001}.

\section{Analysis}
The STAR detector~\cite{STAR}, due to its azimuthal symmetry and large
acceptance, is ideally suited for measuring elliptic flow. For
collisions in its center, the STAR TPC measures the tracks of charged
particles in the pseudo-rapidity range~$|\eta| < 1.8$ with $2\pi$
azimuthal coverage. A scintillator barrel, the Central Trigger Barrel
(CTB), surrounding the TPC that measures the charged particle
multiplicity within $|\eta| < 1$ was used for a central trigger. Two
Zero-Degree Calorimeters~\cite{ZDC} at both ends of the TPC in
coincidence provided a minimum bias trigger.

\begin{figure}[htb]
\centering\mbox{
\includegraphics[width=0.7\textwidth]{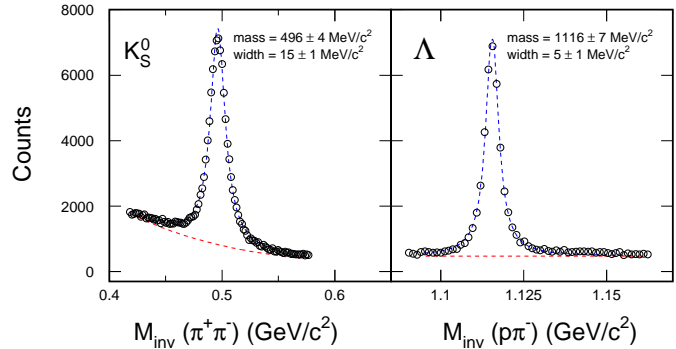}}
\caption{Invariant mass distributions for $\pi^{+}\pi^{-}$ showing an
enhancement at the $K_S^0$ mass (left panel) and
$p\pi^{-}(\overline{p}\pi^{+})$ showing an enhancement at the $\Lambda
(\overline{\Lambda})$ mass (right panel). Fitting results are shown as
dashed lines in the figure. }
\label{fig1}
\end{figure}

The masses and kinematic properties of both $K_S^0 \rightarrow
\pi^+~+~\pi^-$ and $\Lambda(\overline{\Lambda}) \rightarrow p~+~\pi^-
(\overline{p}~+~\pi^+)$, are reconstructed via their decay topologies
in the
TPC~\cite{Wieman:1997rs,Betts:1997dr,Klein:1996fi}. Figure~\ref{fig1}
shows the invariant mass distributions for a $\pi^{+}\pi^{-}$ mass
hypothesis and a $p\pi^- (\overline{p}\pi^+)$ hypothesis. The
background is dominated by combinatoric counts and the observed
masses, $496 \pm 4$ MeV/$c^2$ and $1116 \pm 7$~MeV/$c^2$, are
consistent with values listed in the PDG~\cite{pdg} for $K^0_S$ and
$\Lambda$ respectively. The $K^0_S$ and $\Lambda$ particles used in
the $v_2$ analysis are from the kinematic region of $0.20 \le p_t \le
3$~GeV/c and $|y| \le 1.0$. To compensate for limited statistics,
$\Lambda$ and $\overline{\Lambda}$ are summed together. To reduce the
combinatoric background, for $K^0_S$, pion-like tracks are required to
have a distance of closest approach dca~$>~1.0$~cm, while for
$\Lambda$, the pion-like tracks have a dca~$>~1.5$~cm and the
proton-like tracks have a dca~$>~0.8$~cm. Tracks are determined to be
either proton-like or pion-like based on their energy loss (dE/dx) in the TPC
gas. The yield from the enhancement in the invariant mass peaks in
each $\phi$, $p_t$ bin is used to evaluate $v_2 = \langle
cos(2\phi)\rangle$ as a function of $p_t$.  This method enables us to
measure identified particle flow beyond the $p_t$ range where dE/dx
particle identification fails.

The event plane, an experimental estimator of the true reaction
plane~\cite{art98}, is calculated from the azimuthal distribution of
tracks using cuts similar to those used in
reference~\cite{starflow1}. To avoid auto-correlations, only tracks
excluded from the neutral vertex reconstruction are used in the event
plane calculation. The observed $v_2$ is corrected to account for the
imperfect event plane resolution estimated using the method of
subevents described previously~\cite{art98}. The maximum resolution
correction factor for the $K^0_S$ and $\Lambda$ analysis is found to
be $0.681 \pm 0.004$ and $0.582 \pm 0.007$ respectively and is reached
in the centrality corresponding to 25 -- 35\% of the measured cross
section, where the relative multiplicity distribution is used to
estimate the event centrality as in reference~\cite{starflow1}.

For this analysis, three sources contribute to systematic errors in
the measured anisotropy parameters: (1) particle identification; (2)
background subtraction; (3) non-reaction plane related correlations
contributing to $v_2$ such as resonance decays or Coulomb and
Bose-Einstein correlations~\cite{olliplb477,olliprc62}. The first two
sources are estimated by examining the variation in $v_2$ after
changing several track, event and neutral vertex cuts and are found
to contribute an error of less than~$\pm$ 0.005 to $v_2$. A previous
study used the correlation of event plane angles from subevents to
estimate the magnitude of non-reaction plane related
correlations~\cite{starflow2}. That analysis showed that these effects
which always act to increase the measured value of $v_2$ above its
true value typically contribute a systematic error to $v_2$ of -0.005,
but that the magnitude is larger in the more peripheral events where the
error increases to about -0.035 for the centrality corresponding to 58 -- 85\%
of the measured cross section.

\section{Results}

\begin{figure}[htb]
\centering\mbox{
\includegraphics[width=0.75\textwidth]{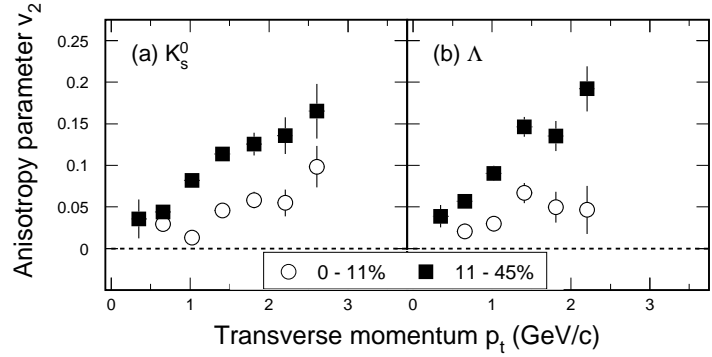}}
\caption{Elliptic flow, $v_2(p_t)$ for $K^0_S$ and $\Lambda$ for central (0 --
11\%) and mid-central (11 -- 48\%) collisions. }
\label{fig2}
\end{figure}

The centrality dependence of $v_2$ as a function of transverse
momentum calculated from 201 thousand minimum bias and 180 thousand
central events is shown in figure~\ref{fig2}. The two particles show a
similar $p_t$ dependence in the respective centralities with more flow
in the more peripheral collisions.  This is similar to observations
made previously at the same energy~\cite{starflow2} where the
agreement with hydrodynamic calculations in the lower $p_t$ region was
interpreted as evidence for early local thermal equilibrium in all but the
most peripheral events (45 -- 85\% of the measured cross section).

\begin{figure}[htbp]
 \resizebox{.53\textwidth}{!}{\includegraphics{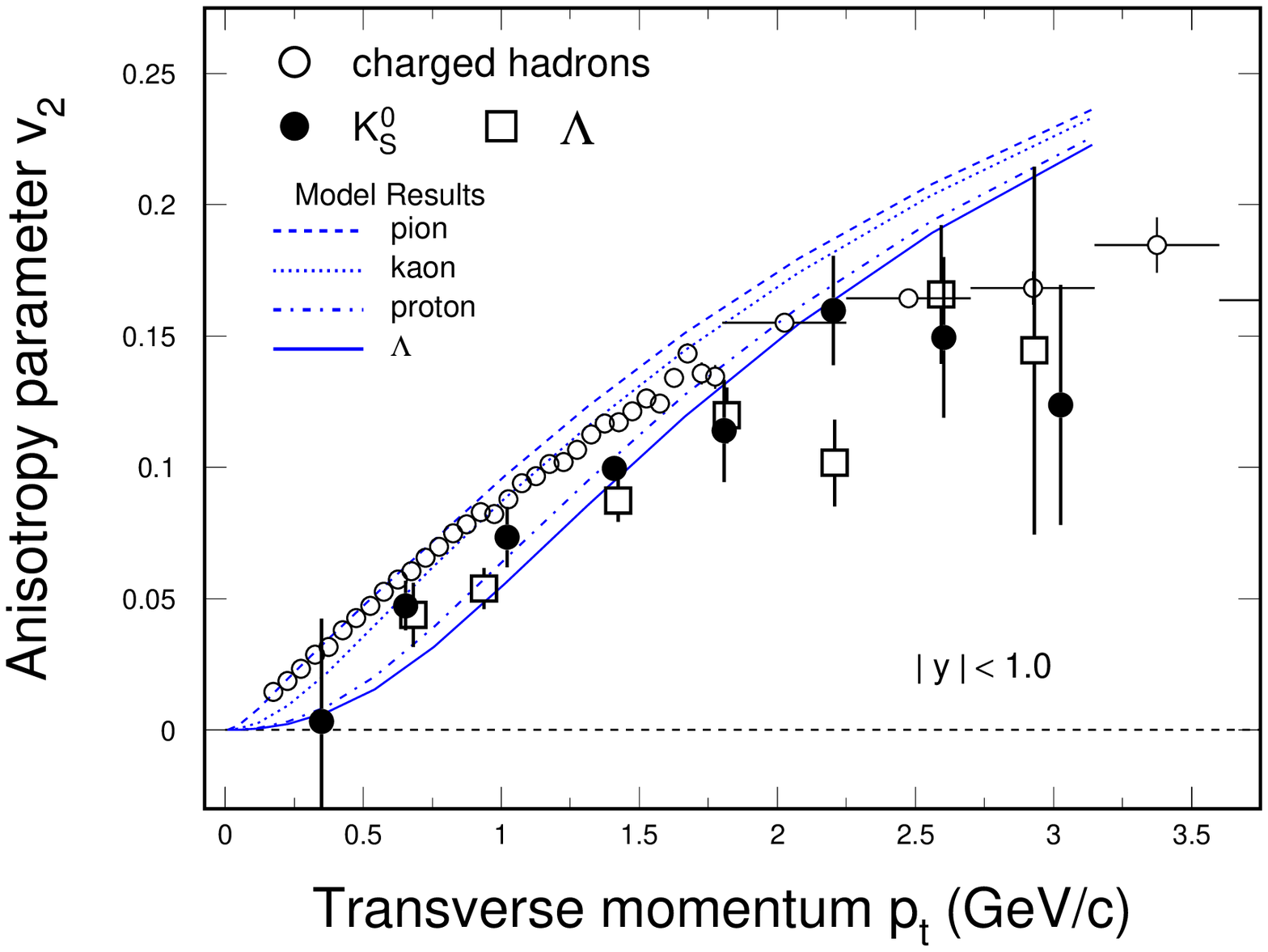}}
 \resizebox{.44\textwidth}{!}{\includegraphics{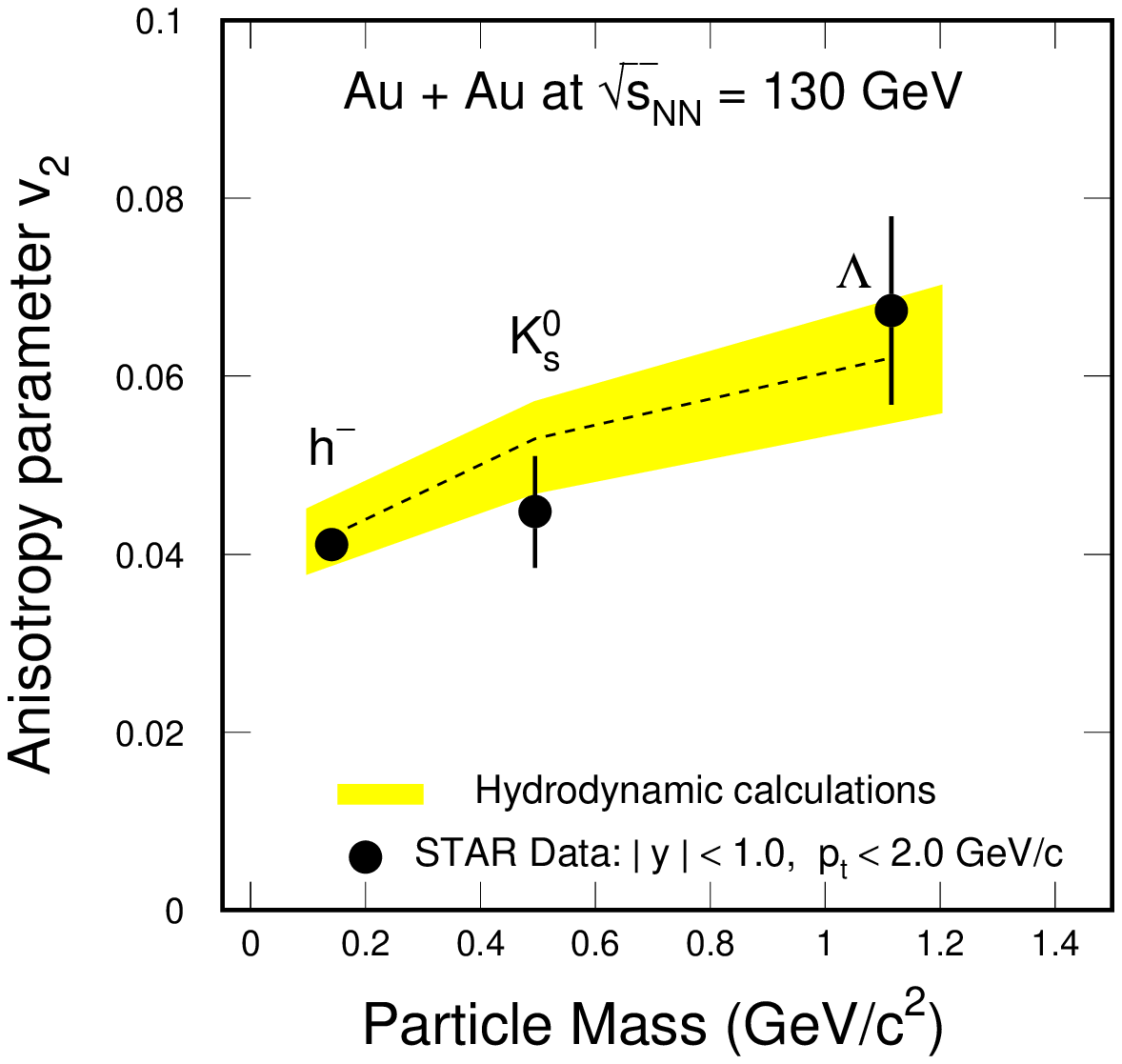}}
\caption{Elliptic flow, $v_2$ for $K^0_S$ and $\Lambda$ as a function
of $p_t$ from minimum bias Au+Au collisions compared to results from
hydrodynamic model calculations and $v_2$ of negatively charged
particles~\cite{qm2001} (left). Integrated azimuthal anisotropy
parameters $v_2$ as a function of particle mass with a gray-band and
central line indicating hydrodynamic model results~\cite{pasi01}
(right).}
\label{fig3}
\end{figure}

In figure~\ref{fig3} (left) we plot $v_2(p_t)$ for $K_S^0$ and
$\Lambda$ from minimum bias collisions with results from hydrodynamic
model calculations~\cite{pasi01} and $v_2(p_t)$ for negatively charged
particles~\cite{qm2001}. The $K_S^0$ results are in agreement with the
$v_2$ for $K^{\pm}$ in the $p_t$ range they share
($300~\le~p_t~\le~700$~MeV/c)~\cite{starflow2}.  We observe that $v_2$
for both strange particles increases as a function of $p_t$ similar to
the hydrodynamic model prediction, up to about 1.5~GeV/c. In the
higher $p_t$ region however ($p_t \ge 2$~GeV/c), the values of $v_2$
seem to be saturated. It has been suggested~\cite{gyulassy01} that the
shape and height of $v_2$ above 2 -- 3~GeV/c is related to energy loss
in an early, high parton-density stage of the evolution.

The $p_t$ integrated $v_2$ from minimum bias collisions for $K_S^0$,
$\Lambda$ and negatively charged particles are shown in
figure~\ref{fig3} (right). The integrated values of $v_2$ are
calculated by parameterizing the yield with the inverse slope
parameter of exponential fits to the $K^0_S$ or $\Lambda$ transverse
mass distributions and are dominated by the region near the particles
mean $p_t$. The relatively larger $v_2$ of $\Lambda$ reflects the
higher mean~$p_t$ of the $\Lambda$ compared to the
$K^0_S$. Hydrodynamic model calculations~\cite{pasi01}, shown as a
gray-band and central line, are, within errors, in agreement with this
result. The width of the gray-band indicates the uncertainty of the
model calculation, mostly due to the choice of the freeze-out
conditions. The increase of $v_2$ with particle mass in
figure~\ref{fig3} points to a significant commonality in velocities
between particles of different masses that is perhaps, established
early in the collision.  The nature of the particles during this
process however, whether parton or hadron, and the degree of
thermalization remains unclear.

\section{Summary}
We have reported the first measurement of $v_2$ for $K_S^0$ and
$\Lambda$ from Au + Au collisions at $\sqrt{s_{_{NN}}}=130$~GeV and
the first measurement, at this energy, of $v_2$ above
$p_t~\sim~0.8$~GeV/c for any identified particle.  For both particles
more flow is seen in mid-central than in central collisions.  The
integrated values of $v_2$ show a mass dependence consistent with the
development of a common velocity, a feature of hydrodynamic models,
where local thermalization is assumed. In $v_2(p_t)$ however, we see a
strong $p_t$ dependence only up to $p_t~\sim$~2.0~GeV/c where $v_2$
seems to saturate, suggesting that the hydrodynamic picture is
incomplete for particle production above $p_t =$~1.5 -- 2.0~GeV/c.


\begin{theacknowledgments}
We thank P. Huovinen for hydrodynamic model calculations and the RHIC
Operations Group at Brookhaven National Laboratory for their support
and for providing collisions for the experiment. This work was
vsupported by the Division of Nuclear Physics and the Division of High
Energy Physics of the Office Science of the U.S. Department of Energy,
the United States National Science Foundation, the Bundesministerium
f\"{u}r Bildung und Forschung of Germany, the Institut National de la
Physique Nucleaire et de la Physique des Particules of France, the
United Kingdom Engineering and Physical Sciences Research Council, and
the Russian Ministry of Science and Technology.
\end{theacknowledgments}

\bibliographystyle{aipproc}

\bibliography{sorensen_v0v2}

\end{document}